\documentclass[fleqn,10pt]{wlscirep}
\title{Switchable plasmonic routers controlled by external magnetic fields  by using magneto-plasmonic waveguides}

\author[1]{Kum-Song Ho}
\author[1,*]{Song-Jin Im}
\author[1]{Ji-Song Pae}
\author[1]{Chol-Song Ri}
\author[1]{Yong-Ha Han}
\author[2,+]{Joachim Herrmann}
\affil[1]{Department of Physics, Kim Il Sung University, Pyongyang, Democratic People's Republic of Korea}
\affil[2]{Max-Born-Institute for Nonlinear Optics and Short Pulse Spectroscopy, Max-Born-Str. 2a,
D-12489 Berlin, Germany}
\affil[*]{ryongnam31@yahoo.com or sj.im@ryongnamsan.edu.kp}
\affil[+]{jherrman@mbi-berlin.de}

\begin{abstract}
We analytically and numerically investigate magneto-plasmons in metal films surrounded by a ferromagnetic dielectric. In such waveguide using a  metal film with a thickness exceeding the Skin depth,  an external magnetic field in the transverse direction can induce a significant spatial asymmetry of mode distribution. Superposition of the odd and the even asymmetric modes over a distance leads to a concentration of the energy on one interface which is switched to the other interface by the magnetic field reversal. The requested magnitude of magnetization is exponentially reduced with the increase of the metal film thickness. Based on this phenomenon, we propose a waveguide-integrated magnetically controlled switchable plasmonic routers with 99-\%-high contrast within the optical bandwidth of tens of THz. This configuration can also operate as a magneto-plasmonic modulator.
\end{abstract}
\begin{document}

\flushbottom
\maketitle

\thispagestyle{empty}

\section*{Introduction}

Plasmonic  waveguides  open  the possibility to develop dramatically miniaturized optical devices  and provide a promising route for the development of next-generation integrated nanophotonic circuits \cite{Fang2015}. Key elements of nanophotonic circuits are plasmonic modulators, controllable plasmonic splitters and routers. Plasmonic modulators has been studied  using diverse principles including the Pockels effect \cite{Schildkraut1988,	Melikyan2014,	Haffner2015}, the thermo-optical  effect \cite{Nikolajsen2004,	Gosciniak2013}, and gain-assisted cavity coupling \cite{Yu2008,	Cai2009,	Im2016}. Beam splitters and demultiplexers  have  been realized by using nanoparticle chains, nanowires, nanogaps and nanogratings, but active control of the splitting is absent in these types of devices. Up to our knowledge, there exist only few reports with the realization of switchable plasmon routers. Branched silver nanowires \cite{Fang2010, Wei2015} have been used for surface plasmon polariton (SPP) switching between branched silver nanowires by controlling the polarization of the input light. Besides a periodic  slanted gold mushroom array has been developed  as a switchable SPP splitters \cite{Shen2016}.

In recent years the combination of plasmonics with magnetism become an active topic of research to achieve new functionalities in nanosystems \cite{Armelles2013}.
In applications for plasmonic modulators, the magneto-optical effect has the advantage compared with other proposed principles such as the  electro-optical effect and thermo-optical effect that the switching can be much faster. The generation of magnetic fields by integrated electronic circuits can easily reach speeds in the GHz regime.
Most of the proposed configurations for ultrafast magneto-plasmonic modulators in the visible and near-infrared range are based on the magnetically-induced wavenumber change of SPPs \cite{Sepulveda2006,	Khurgin2006}. The performance of magneto-plasmonic modulators generally suffer from the small wavenumber changes. The introduction of photonic crystals \cite{Yuu2008,	Kuzmiak2012} could provide larger wavenumber changes at the expense of reduced transmission bandwidth determined by the photonic bandgap. The configuration using the plasmon-enhanced transverse magneto-optical Kerr effect in nanopatterned gold/garnet structures \cite{Belotelov2011,	Kreilkamp2013,	Belotelov2013}, being relatively bulky and not fully waveguide-integrated, was experimentally demonstrated to be resonantly sensitive to the magnetically-induced changes of SPP wavenumber and allow a modulation contrast of 50\%. A more compact configuration based on a magneto-plasmonic interferometer engraved in a hybrid ferromagnetic structure was successfully developed \cite{Temnov2010,	Armelles2013,	Temnov2016,	Martin2010,	Ferreiro2011,	Temnov2012}. Its modulation contrast was on the level of 2\% at a waveguide length of 20 $\mu$m \cite{Temnov2010}, however this can be increased to 12\% \cite{Armelles2013} by adding a higher refractive index dielectric layer on top of the interferometer and by optimizing the ferromagnet layer.

In this paper, we present a new conception for a switchable plasmonic router based on the magnetically-induced spatial asymmetry of planar SPP modes in an external magnetic field. Asymmetric guiding modes are generally supported by waveguides with asymmetric cross sections. The asymmetry can also be achieved through active approaches, for example, by using power-dependent nonlinear dielectrics \cite{Salgueiro2010,	Arthur2008} or by introducing an external magnetic field \cite{Hu2012,	Rukhlenko2012,
Im2017}. We analytically and numerically study the magnetically-induced mode asymmetry in a metal film surrounded by a ferromagnetic dielectric. An experimentally achievable magnetization can induce a significant mode asymmetry for a metal film with thickness exceeding the Skin depth. For a certain degree of mode asymmetry, a superposition of the even and the odd modes over a distance leads to a high-contrast-asymmetric splitting within a broad optical bandwidth. This configuration can operate as a waveguide-integrated switchable magneto-plasmonic router with high contrast and broad optical bandwidth controlled by an external magnetic field.

\section*{Results and Discussion}
\textbf{Magnetically-induced mode asymmetry}

\begin{figure}
\centering
\includegraphics[width=0.6\textwidth]{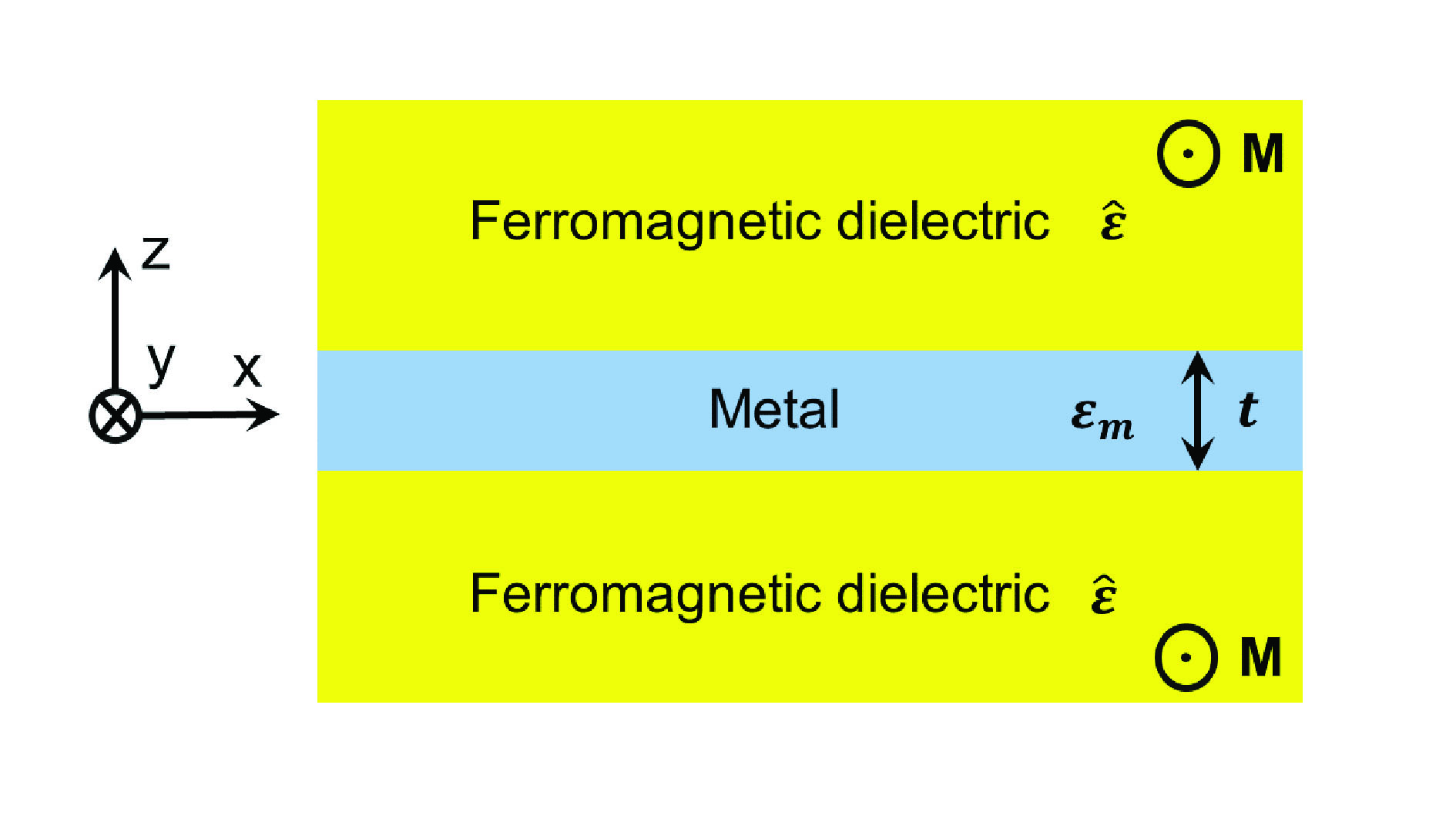}
\caption{Scheme of a magneto-plasmonic waveguide with an external magnetic field inducing a magnetization \textbf{M} in the transverse y-direction.}
\label{fig1}
\end{figure}

We study analytically and numerically by solving the Maxwell
equations the magnetically-induced mode asymmetry in a metal film surrounded by a ferromagnetic dielectric under an external magnetic field inducing a magnetization  $\textbf{M}=M \hat{y}$, where $ \hat{y}$ is the unit vector in y-direction, as shown in Fig. \ref{fig1}. The higher absorption in ferromagnetic metals results in shorter propagation lengths of SPPs and makes these materials unfavorable for many applications. Ferromagnetic dielectric materials offer the advantage of much lower loss which makes a noble metal film surrounded by a ferromagnetic dielectric attractive for diverse applications. Excitation of SPP modes in diverse metal film
structures can be achieved by the Kretschman configuration or
a grating (see e.g. Ref. \cite{Berini_2009}). For a sufficiently large thickness of the surrounding dielectric layer, the SPP modes are not significantly influenced by the substrate material. We assume that the metal layer is surrounded by semi-infinite dielectric layers.

The permittivity tensor $\hat \varepsilon$ of the ferromagnetic dielectric can be expressed as follows. 
\begin{eqnarray}
\hat \varepsilon  = \left( {\begin{array}{*{20}{c}}
{{\varepsilon _d}}&0&{ig}\\
0&{{\varepsilon _d}}&0\\
{ - ig}&0&{{\varepsilon _d}}
\end{array}} \right)
\label{eq1}
\end{eqnarray}
Here $g$ is defined as
\begin{eqnarray}
g=aM,
\label{eq2}
\end{eqnarray}
where $a$ is the magneto-optical susceptibility.
For the fundamental TM mode, we can get the following dispersion relation from the Maxwell equation and boundary conditions.
\begin{eqnarray}
{e^{ - 2{k_m}t}} = \frac{{{q_m} + {q_{ - g}}}}{{{q_m} - {q_{ - g}}}}\frac{{{q_m} + {q_g}}}{{{q_m} - {q_g}}}
\label{eq3}
\end{eqnarray}
where ${q_m}={k_m}/{\varepsilon _m}$, ${q_g}=\left( {{\varepsilon _d}{k_d} + g\beta } \right)/\left( {\varepsilon _d^2 - {g^2}} \right)$, $k_m^2 = {\beta ^2} - k_0^2{\varepsilon _m}$, $k_d^2 = {\beta ^2} - k_0^2(\varepsilon _d^2 - {g^2})/{\varepsilon _d}$, $t$ is the  thickness of the metal film, $\beta$ is the wavenumber, and ${\varepsilon _m}$ and ${\varepsilon _d}$ are the permittivity of the metal and the ferromagnetic dielectric, respectively. In the absence of the external magnetic field $g=0$,  the dispersion relation Eq.~(\ref{eq3}) is simplified to  $\tanh ({k_m}t/2) =  - ({k_d}{\varepsilon _m})/({k_m}{\varepsilon _d})$ for the odd mode and  $\tanh ({k_m}t/2) =  -( {k_m}{\varepsilon _d})/({k_d}{\varepsilon _m})$ for the even mode \cite{ Maier2007}.

For a metal film thickness larger than the Skin depth $\delta  = 1/2{k_m}$, in the first order of approximation with respect to $|(\beta-\beta_0)/\beta_0|$,  $\beta ^{e(o)}$  is given by
\begin{eqnarray}
{\beta ^{e(o)}}  \approx \beta_0  +(-) \sqrt{{\left( \Delta \beta_0 \right)}^2+{{\left( \Delta \beta_g \right)}^2}},
\label{eq4}
\end{eqnarray}

\begin{eqnarray}
{ \Delta \beta_0}= \frac{2{{\beta_0}}}{{1 - \varepsilon _d^2/\varepsilon _m^2}}\left( { - {\varepsilon _d}/{\varepsilon _m}} \right){\exp{(- t/2\delta)}},  \qquad  { \Delta \beta_g}=\frac{{\beta_0}}{{1 - \varepsilon _d^2/\varepsilon _m^2}}\frac{{g}}{\sqrt{-{\varepsilon _m} {\varepsilon _d}}}.
\label{eq5}
\end{eqnarray}
where $\beta_0=k_0\sqrt{\left(\varepsilon _d\varepsilon _m\right)/\left(\varepsilon _d+\varepsilon _m\right)}$.  The superscripts $o$ and $e$ represent the odd and the even modes, respectively.
We note that ${ \Delta \beta_g}$ is in accordance with Eq.~(11) of Ref.~\cite{Im2017_1} and Eq.~(4) of Ref.~\cite{Belotelov2011} which describes the magnetically-induced wavenumber change by a single interface between a metal and a ferromagnetic dielectric. If the thickness $t$ is infinite,  the dispersion relation Eq.~(\ref{eq3}) is simplified to $\beta=\beta_0(1\pm g/\sqrt{-({\varepsilon _{m}}+{\varepsilon _{d}} 
)}/(1-\varepsilon _{d}^{2}/\varepsilon _{m}^{2})$ in agreement with Eq. (4) of Ref. \cite{ Belotelov2011}.

We define the degree of the mode asymmetry as follows.
\begin{eqnarray}
{m^{o(e)}} = \frac{{H_y^{o(e)}\left( {z =  - t/2} \right)}}{{H_y^{o(e)}\left( {z = t/2} \right)}},
\label{eq6}
\end{eqnarray}
where $H_y^{o(e)}\left( {z =  - t/2} \right)$ and $H_y^{o(e)}\left( {z = t/2} \right)$ are the magnetic field component of the odd (even) mode at the bottom and the top interfaces of Fig. \ref{fig1}, respectively. 

From the boundary condition  and dispersion relation Eq.~(\ref{eq3}), we can express the degree of the mode asymmetry as follows. 
\begin{eqnarray}
m^{o(e)} = \frac{{{q_m^{o(e)}} + {q_g^{o(e)}}}}{{{q_m^{o(e)}} - {q_{ - g}^{o(e)}}}}{e^{t/2\delta }}
\label{eq7}
\end{eqnarray}
In the absence of an external magnetic field, the intensity distributions for both modes are symmetric, ${m^{o}} = 1$ and ${m^{e}} =  -  1$ but an external magnetic field induces an asymmetry of the mode distribution. By using  Eq.~(\ref{eq4}), in the first order of  $|(\beta-\beta_0)/\beta_0|$, Eq.~(\ref{eq7}) is approximated as follows.
\begin{subequations}\label{eq8}
\begin{align}
{m^{o}} \approx \sqrt {1 + (g/g_0)^2}  - g/g_0, \\
{m^{e}} \approx   -  \sqrt {1 + (g/g_0)^2}  - g/g_0,
\end{align}
\end{subequations}
\begin{eqnarray}
{g_0} = 2{\varepsilon _d}{\left( { - {\varepsilon _d}/{\varepsilon _m}} \right)^{1/2}}\exp{ (- t/2\delta)}.
\label{eq9}
\end{eqnarray}
Here $g_0$ characterize a gyration requested to induce a significant asymmetry of ${m^{o}}=\sqrt{2}-1$ (${m^{e}}=-\sqrt{2}-1$).  Eq.~(\ref{eq9}) shows that the characteristic gyration $g_0$ is exponentially reduced with the increase of the metal film thickness $t$. For a metal film thickness exceeding the Skin depth, a significant asymmetry can be achieved by introducing a small value of $g$ or a corresponding external magnetic field.
\begin{figure}
\centering
\includegraphics[width=0.6\textwidth]{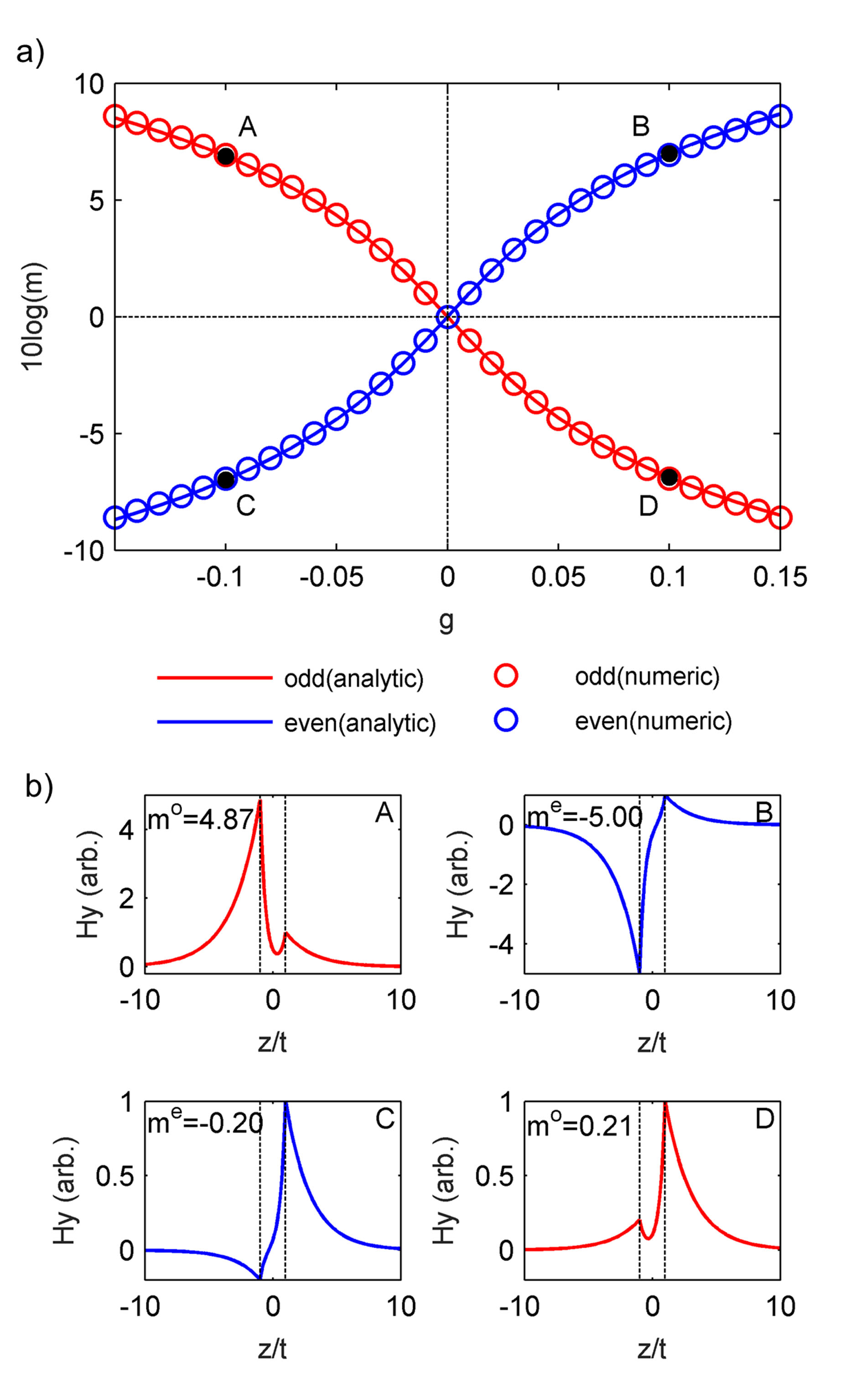}
\caption{Magnetically-induced mode asymmetry in the magneto-plasmonic waveguide, as shown in Fig. \ref{fig1}, with $t = 100$ nm  at  $\lambda  = 800$ nm. The solid curves are calculated by the formula (\ref{eq8}) and the circles by numerical solutions. a) The symmetry degree $m$  for the odd (red) and the even (blue) modes versus $g$. b) The magnetic field profiles  ${H_y}\left( z \right)$ for the points A, B, C and D in a). }
\label{fig2}
\end{figure} 
Fig.~\ref{fig2} shows the asymmetry $m$ for the odd (red curves) and the even (blue curves) modes in the magneto-plasmonic waveguide, as shown in Fig. \ref{fig1}, with $t = 100$ nm  at $\lambda  = 800$ nm  in dependence on $g$. The diagonal permittivity of the ferromagnetic dielectric is assumed to be ${\varepsilon _d} = 6.25$. The experimental data for the permittivity of silver \cite{Johnson1972} are used as ${\varepsilon _m}$. The asymmetry degree (Fig. \ref{fig2}a)) calculated from the formula (\ref{eq8}) (solid curves) well agree with the results from the numerical solution of Eq.~(\ref{eq3}) (circles). The numerically calculated mode profiles for the points A, B, C and D of Fig. \ref{fig2}a) are shown in  Fig. \ref{fig2}b).

\textbf{High-contrast magneto-plasmonic modulation}

\begin{figure}
\centering
\includegraphics[width=0.6\textwidth]{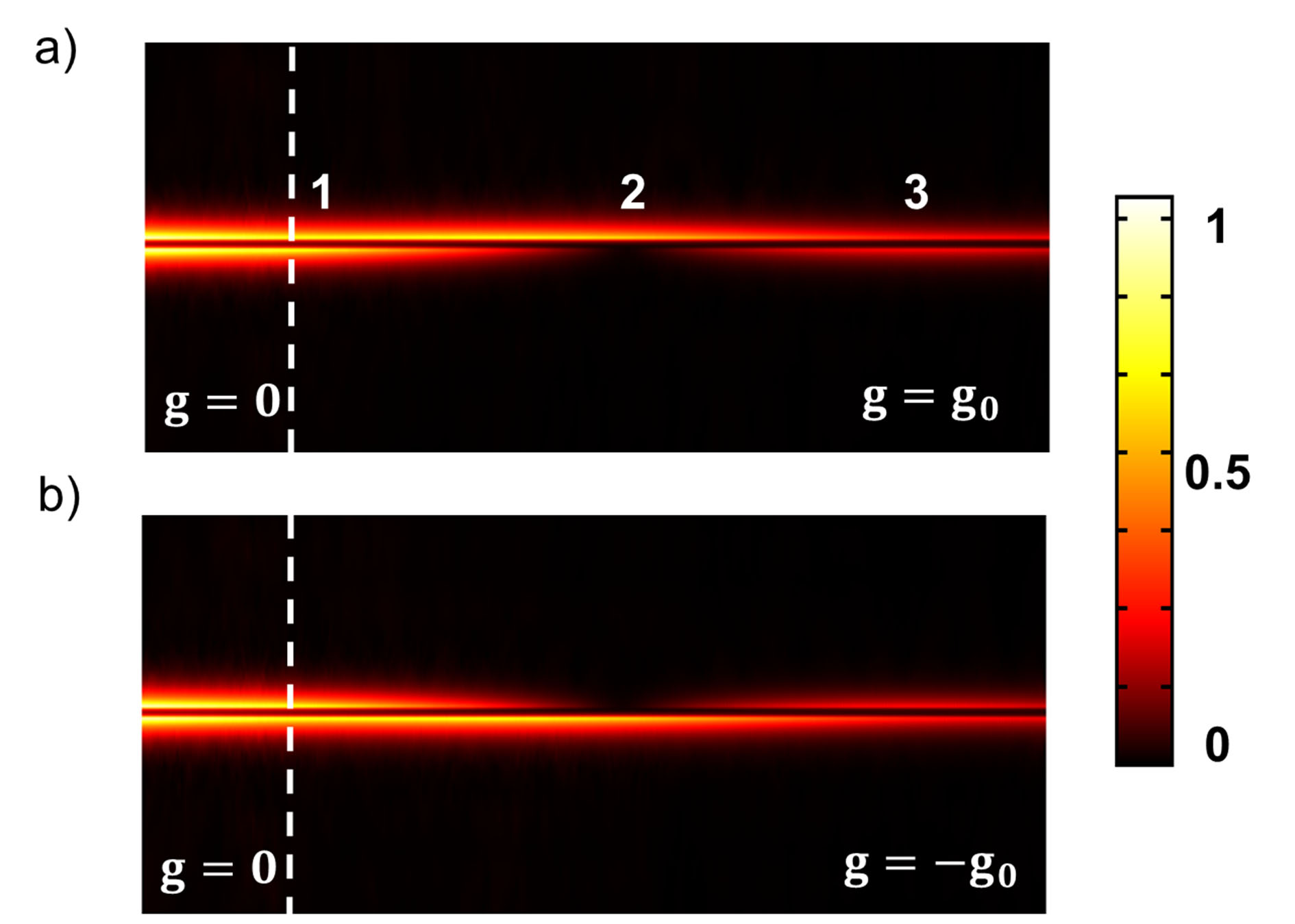}
\caption{Numerical results of the distribution of ${\left| {{H_y}} \right|^2}$  in the magneto-plasmonic waveguide. The left and the right parts to the white dashed line are in the absence and the presence of the external magnetic fields inducing $g = \pm{g_0}$, respectively. The value of  ${g_0}$, 0.042 is calculated by Eq.~(\ref{eq9}). The symmetric SPP mode at  $\lambda  = 800$ nm is incident to the left port of the structure. Other parameters are same as in Fig. \ref{fig2}.}
\label{fig3}
\end{figure}

Based on the above results of the magnetically-induced mode asymmetry, a configuration for a high-contrast magnetically controlled switchable plasmonic router is suggested. A coupled plasmonic waveguide without external magnetic field is added to the left side of the magneto-plasmonic waveguide structure, as shown in Fig. \ref{fig3}. The long-range SPP mode (odd mode), which is incident to the left port, is presented by a superposition of the odd and the even modes in the magneto-plasmonic waveguide.
\begin{eqnarray}
{H_y}(x,z) = {c_1}H_y^o(z)\exp (i{\beta ^o}x) + {c_2}H_y^e(z)\exp (i{\beta ^e}x)
\label{eq10}
\end{eqnarray}

The coefficients $c_1$ and $c_2$ can be obtained from the continuity of field distribution at the interface between the non-magnetic and the magneto-plasmonic waveguides, shown by the vertical white-dashed-line in  Fig. \ref{fig3}. While the magneto-plasmon propagates along the waveguide, the field distribution during propagation varies due to the difference in wavenumber between the odd and the even modes, ${\beta ^e} - {\beta ^o}$. By using Eqs.~ (\ref{eq7}) and (\ref{eq10}), the intensity ratio
\begin{eqnarray}
R(x) = 10\log {\left| {\frac{{H_y}\left( {x,z =  - t/2} \right)}{{H_y}\left( {x,z = t/2} \right)}} \right|^2}
\label{eq11}
\end{eqnarray}
can be expressed as follows.
\begin{eqnarray}
R\left( x \right) = 10 \log{\left| {\frac{{{m^o}^2 + {m^o} - \left( {{m^o} - 1} \right)\exp \left[ {i({\beta ^e} - {\beta ^o}) x} \right]}}{{\left( {{m^o}^2 - {m^o}} \right)\exp \left[ {i({\beta ^e} - {\beta ^o})x} \right] + {m^o} + 1}}} \right|^2}
\label{eq12}
\end{eqnarray}
At the half period ${L_{\rm{0}}}=\pi / ({\beta ^e} - {\beta ^o})$, it reaches its extreme value
\begin{eqnarray}
R({L_{\rm{0}}}) = 10\log {\left( {\frac{{{m^o}^2 + 2{m^o} - 1}}{{ - {m^o}^2 + 2{m^o} + 1}}} \right)^2}.
\label{eq13}
\end{eqnarray}
Here we assumed $\texttt{Im}\beta\ll \texttt{Re}\beta$.
By substituting Eq.~(\ref{eq9}) in Eq.~(\ref{eq13}), we obtain
\begin{eqnarray}
R({L_{\rm{0}}}) \approx 10\log {\left( {\frac{{g - {g_0}}}{{g + {g_0}}}} \right)^2}.
\label{eq14}
\end{eqnarray}
From Eq.~(\ref{eq14}), one can see that the ratio $R$ becomes extremely high at the characteristic gyration  $g =  \pm {g_0}$.  The magnitude of magnetization for a high contrast exponentially decreases with the increase of the metal film thickness $t$.

Let's investigate the half period  ${L_{\rm{0}}}$ for the case of a high contrast. At the high-contrast points $g =  \pm {g_0}$, we have
\begin{eqnarray}
{\beta ^e} - {\beta ^o}\approx  \frac{ 4\sqrt 2{{\beta_0}}}{{1 - \varepsilon _d^2/\varepsilon _m^2}}\frac { - {\varepsilon _d}}{\varepsilon _m} \exp (-t/2\delta),
\label{eq15}
\end{eqnarray}
\begin{eqnarray}
{L_{\rm{0}}} \approx \frac{\pi }{{4\sqrt 2 {\beta_0}}}\left( {1 - \varepsilon _d^2/\varepsilon _m^2} \right)\frac { - {\varepsilon _m}}{\varepsilon _d} \exp (t/2\delta).
\label{eq16}
\end{eqnarray}
Fig.~\ref{fig3} shows results of the numerical simulation for the distribution of  ${\left| {{H_y}} \right|^2}$ in the magneto-plasmonic waveguide. The left and the right parts to the white dashed line are in the absence and the presence of the external magnetic fields inducing $g = \pm{g_0}$, respectively. One can see that the field intensity is concentrated on the bottom interface, which is switched to the top interface by the reversal of the magnetic field direction. The distance between the point 1 and the point 2 is coincident with  ${L_0} = 16$ $\mu$m calculated by Eq.~(\ref{eq16}). The field distribution is recovered at the point 3 except a power attenuation due to the ohmic loss in the metal. 

\begin{figure}
\centering
\includegraphics[width=0.6\textwidth]{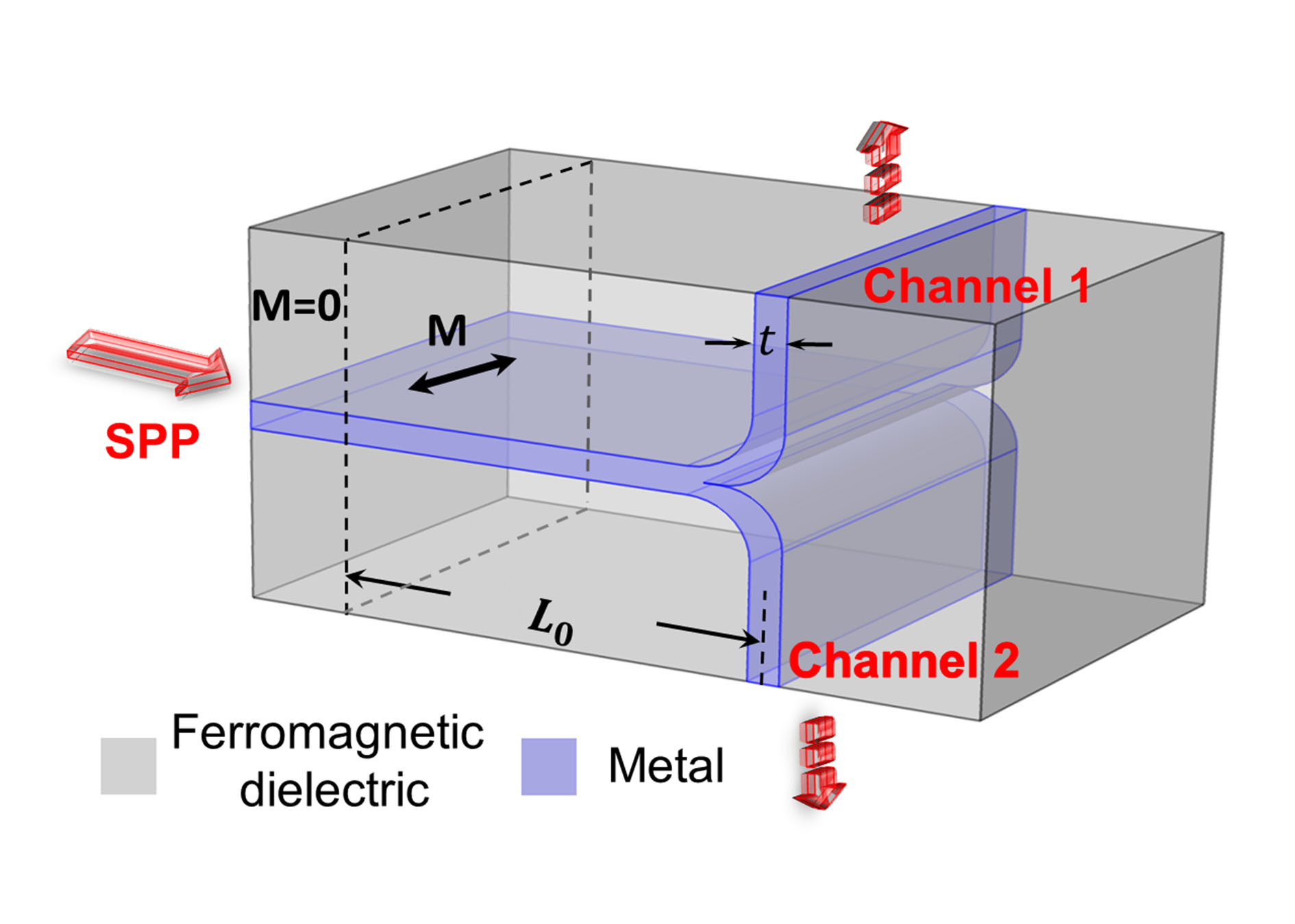}
\caption{A configuration for a switchable plasmonic router. }
\label{fig4}
\end{figure}

A configuration for the magnetically controlled switchable plasmon router is shown in Fig. \ref{fig4}.
The value of ${g_0}$ can be reduced by increasing the metal thickness to a value of 110 nm. ${g_0} = 0.026$ and  ${L_{\rm{0}}}{\rm{ = }}26$ $\mu$m have been calculated by Eqs.~(\ref{eq9}) and (\ref{eq16}). We note that the experimentally observed gyration values of 0.03 or 0.06 for Bi-substituted iron garnet (BIG) have been reported in  Ref. \cite{Yao2015}, in the Table 1 of Ref. \cite{Dutta2017} and in the Methods of Ref. \cite{Davoyan2014}. We assume a bending radius of 300 nm for which the numerical simulation shows a small value of the bending loss of 1.2\%.  For a more realistic simulation we take into account the absorption of 300 dB/cm \cite{Zvezdin1997} in the ferromagnetic dielectric. The calculation using Eq.~(3) predicts ${\texttt{Im}}\beta \approx 5\times10^4$/m including  a contribution of the absorption in the ferromagnetic dielectric, $1\times10^4$/m, which is not ignorable compared to ${\beta ^e} - {\beta ^o}\approx 12\times10^4$/m. The absorption in the metal and the ferromagnetic dielectric results in an insertion loss, however does not influence the high contrast because the intensity reduction due the absorption occur equally in the both channels.

\begin{figure}
\centering
\includegraphics[width=0.6\textwidth]{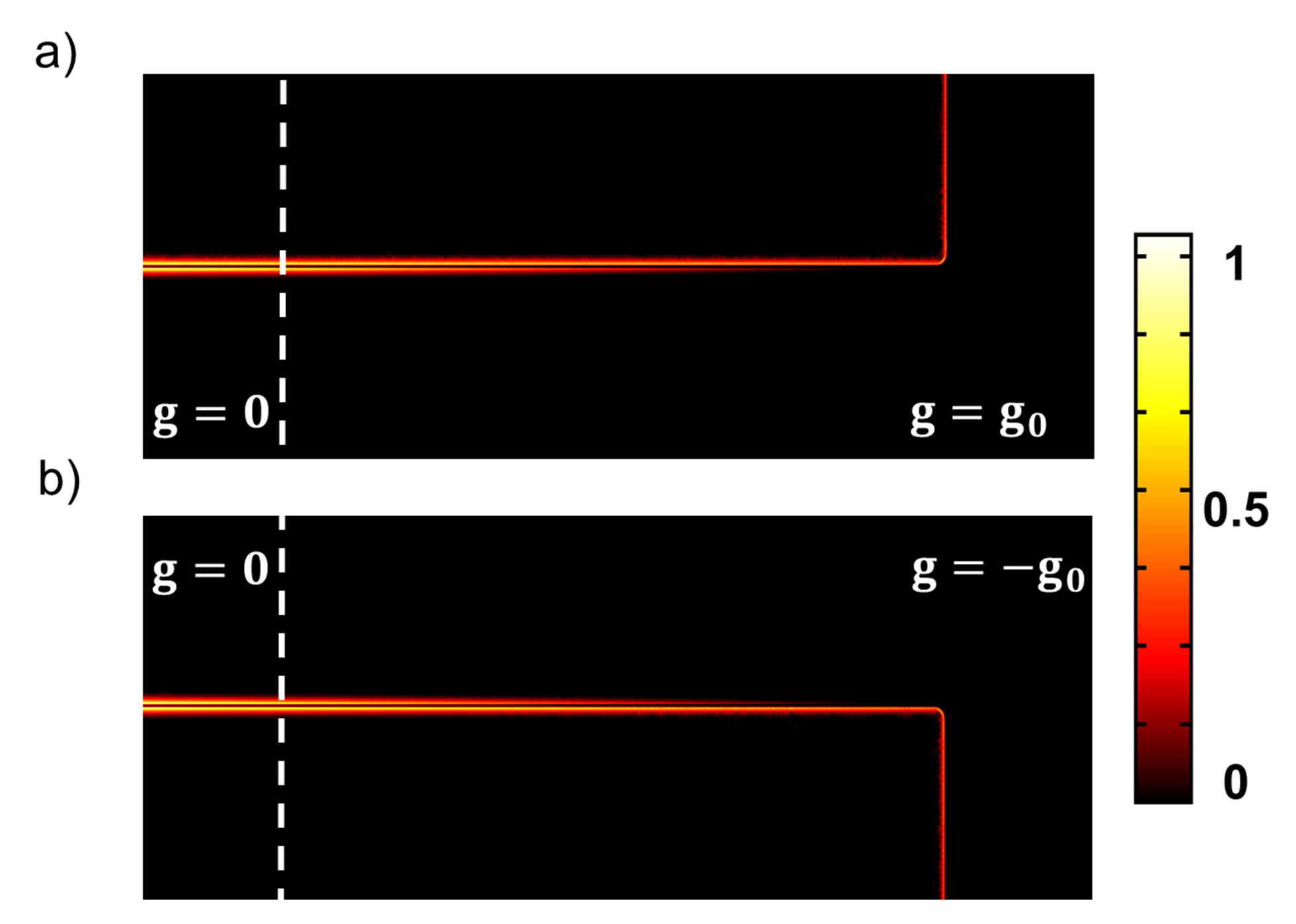}
\caption{Magnetically controlled switchable plasmon router. Numerical results of the distribution of ${\left| {{H_y}} \right|^2}$. Here ${g_0} = 0.026$  and  ${L_{\rm{0}}}{\rm{ = }}26$ $\mu$m are calculated by Eqs.~(\ref{eq9}) and (\ref{eq16}). The metal thickness has been increased to a value of 110 nm. The other parameters are the same as in Fig. \ref{fig3}.}
\label{fig5}
\end{figure}

In Fig.~\ref{fig5} numerical results for the configuration show a high-contrast magneto-plasmonic modulation and channel switching by the magnetic field reversal.

\begin{figure}
\centering
\includegraphics[width=0.6\textwidth]{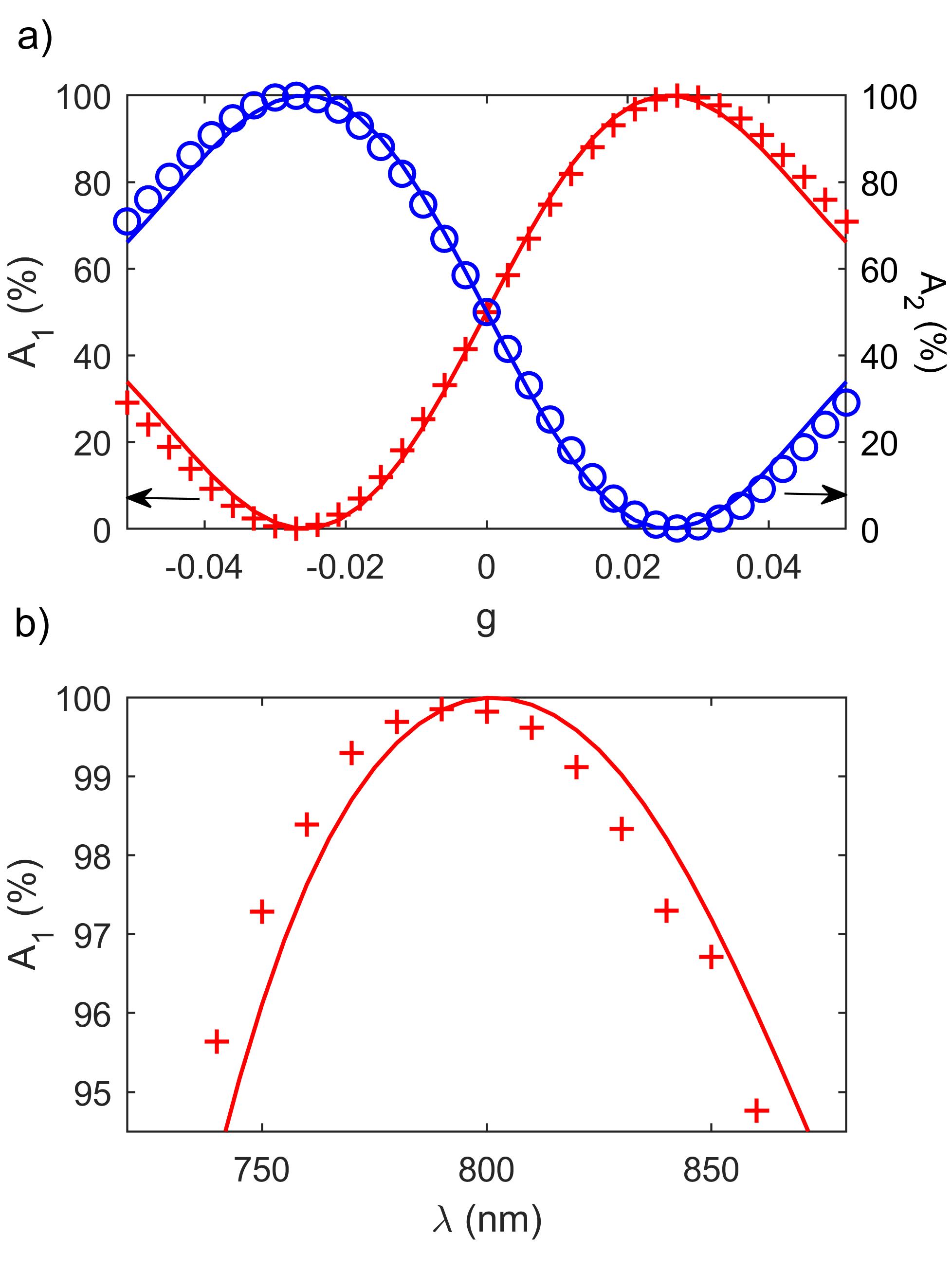}
\caption{Modulation contrast $A_1=P_{ch1}/(P_{ch1}+P_{ch2})$ and $A_2=P_{ch2}/(P_{ch1}+P_{ch2})$ defined by the ratio of powers through the two channels in the configuration in dependence on the external magnetic field  $g$ (a) and the wavelength $\lambda $ (b). Other parameters are the same as in Fig. \ref{fig5}. The red crosses and the blue circles are calculated numerically and the red and the blue curves from the analytical formula.}
\label{fig6}
\end{figure}

We estimate the modulation contrast $A_1=P_{ch1}/(P_{ch1}+P_{ch2})$ and $A_2=P_{ch2}/(P_{ch1}+P_{ch2})$ defined by the ratio of powers through the two channels in the configuration in dependence on $g$ (Fig. \ref{fig6}a)) and the wavelength $\lambda $(Fig. \ref{fig6}b)) for  ${L_{\rm{0}}}{\rm{ = }}26$ $\mu$m. The numerical results (red crosses) well agree with the analytically predicted results by Eq.~(\ref{eq12}) (blue curves). The power splitting ratio is tunable by controlling the external magnetic field (Fig. \ref{fig6}a)). In the vicinity of  $g =  \pm {g_0} =  \pm 0.026$  the power splitting ratio reaches a value larger than 20 dB, which allows a high-contrast magneto-plasmonic modulation.  The 99-\%-high contrast is maintained in a broad wavelength range from 770 nm to 820 nm (Fig. \ref{fig6}b)).

\begin{figure}
\centering
\includegraphics[width=0.6\textwidth]{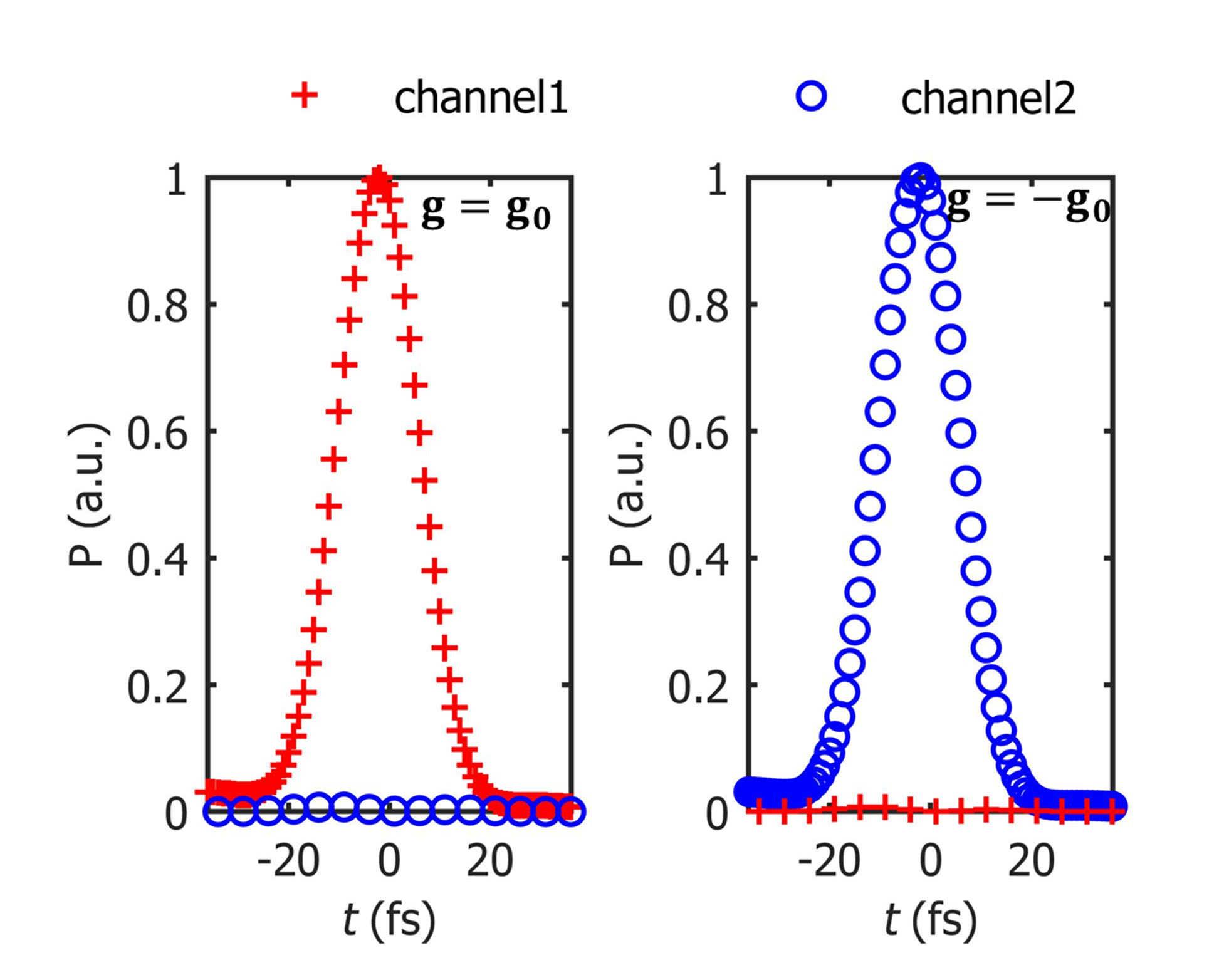}
\caption{Temporal power characteristics of the SPPs in the configuration Fig. \ref{fig4}a). The full-width at half maximum (FWHM) of the pulse is 20 fs and the central wavelength is 800 nm. The red crosses represent the power through channel 1 while the blue circles the power through channel 2.  Other parameters are same as in Fig. \ref{fig5}.}
\label{fig7}
\end{figure}

The broad bandwidth shown in Fig. \ref{fig6}b)  allows a high-contrast modulation of femtosecond pulses. The temporal  power of the SPP pulses in both channels  are shown in Fig. \ref{fig7}. The full-width at half maximum (FWHM) of the pulse is 20 fs and the central wavelength is 800 nm. By the magnetic field reversal, the ultrashort pulse is amplitude-shift keying (ASK) modulated with 99-\%-high contrast.

In conclusion, we derived an analytical formula for the magnetically-induced mode asymmetry in metal films, with a thickness exceeding the Skin depth, surrounded by ferromagnetic dielectrics. In such waveguide, a significant spatial asymmetry of the mode distribution can be induced by an external magnetic field in the transverse direction which is exponentially reduced with the increase of the metal film thickness. Superposition of the odd and the even asymmetric modes over a distance presents complete energy concentration on one interface which is switched to the other interface by the magnetic field reversal. Based on this phenomenon, we proposed a waveguide-integrated magnetically controlled switchable plasmonic router with 99-\%-high contrast within an optical bandwidth of tens of THz. This configuration can also operate as a magneto-plasmonic modulator.

\section*{Methods}

Analytical formulas (\ref{eq4}) and (\ref{eq8}) have been derived in the first order approximation with respect to $|(\beta-\beta_0)/\beta_0|$. Details of the derivation of formula (\ref{eq4})  are presented in the Supplementary information.

Numerical solutions for the mode distribution in Fig. \ref{fig2} \cite{Im2017} have been obtained by solving Eq.~(\ref{eq3}).

Numerical simulations for the plasmonic propagation in Fig. \ref{fig3}, Fig. \ref{fig5} and Fig. \ref{fig6} have been performed by numerically solving the Maxwell equation in the frequency domain \cite{Im2016}, where modeling the spatial electromagnetic distribution has been done using the finite element method (FEM). 

The temporal response to the input pulse was calculated by the superposition of the responses of the Fourier components of the time-depending pulse.

The experimental data for the permittivity spectra of silver \cite{Johnson1972} have been used as the permittivity  ${\varepsilon _m}$ of the metal film in the structure of Fig. \ref{fig1}.

\bibliography{mpm}

\section*{Acknowledgements}

\section*{Author contributions statement}

K.H. and S.I. conceived the idea and performed theoretical calculations. K.H., J.P., C.R and Y.H contributed to numerical simulations. K.H., S.I. and J.H analyzed the results and contributed to preparation of the manuscript.  S.I. and J.H supervised the project.

\section*{Additional information}

 \textbf{Supplementary information} 
accompanies this paper at http://www.nature.com/srep

 \textbf{Competing interests:}
The authors declare no competing interests.

\end{document}